\documentclass{juliacon}
\usepackage{caption}
\usepackage[round]{natbib}  
\begin{document}


\title{SequentialSamplingModels.jl: Simulating and Evaluating Cognitive Models of Response Times in Julia}

\author[1]{Kianté Fernandez}
\author[2]{Dominique Makowski}
\author[3]{Christopher Fisher}
\affil[1]{University of California, Los Angeles}
\affil[2]{School of Psychology, University of Sussex, Brighton, UK}
\affil[3]{Independent Researcher}


\hypersetup{
pdftitle = {My JuliaCon proceeding},
pdfsubject = {JuliaCon 2024 Proceedings},
pdfauthor = {1st author, 2nd author, 3rd author},
pdfkeywords = {cognitive models,response time, sequential sampling models, evidence accumulation, decision-making},
}

\maketitle

\begin{abstract}

Sequential sampling models (SSMs) are a widely used framework describing decision-making as a stochastic, dynamic process of evidence accumulation. SSMs popularity across cognitive science has driven the development of various software packages that lower the barrier for simulating, estimating, and comparing existing SSMs. Here, we present a software tool, SequentialSamplingModels.jl (SSM.jl), designed to make SSM simulations more accessible to Julia users, and to integrate with the Julia ecosystem. We demonstrate the basic use of SSM.jl for simulation, plotting, and Bayesian inference.

\end{abstract}

\section{Introduction}
Sequential sampling models (SSMs) are widely used in cognitive science due to their ability to describe dissociable processes underlying a wide variety of capacities, including memory retrieval, visual perception, and decision making (\cite{forstmann2016sequential}). These models typically describe decision-making as a stochastic, dynamic evidence accumulation process which evolves until evidence for one option reaches an evidence threshold (\cite{smith2000stochastic,Ratcliff2008TheDD,Ratcliff1978ATO}). By doing so, SSMs provide a generative model for response time (RT) distributions of actions made by organisms capturing both the speed of the response and the response itself. Figure 1 illustrates the latent evidence accumulation process for a hypothetical decision between a Margherita pizza and a pineapple pizza. Evidence accumulates stochastically (i.e., randomly) until it reaches either the upper boundary, triggering the selection of the Margherita, or the lower boundary, triggering the selection of the pineapple pizza. While these models hold many promising applications, their implementation and use remain a technical challenge.

The widespread interest and continuous use of SSMs across cognitive science have spurred the development of various software packages that lower the barrier for simulating, estimating, and comparing existing SSMs. These software implementations range widely, from those designed to provide maximum flexibility to end-users (\cite{Shinn2020AFF}) to those that streamline the entire analysis of SSMs (\cite{Fengler2022BeyondDD}). Toolboxes have also been developed across multiple programming languages commonly used for scientific computing, including MATLAB (\cite{vandekerckhove2008diffusion}), Python (\cite{wiecki2013hddm, Shinn2020AFF, Fengler2022BeyondDD, murrow2024pybeam}), and R (\cite{wabersich2014rwiener,stevenson2024emc2, hBayesDM,singmann2018package, hartmann2021partial}).

\begin{figure}
    \centering
    \includegraphics[width=1\linewidth]{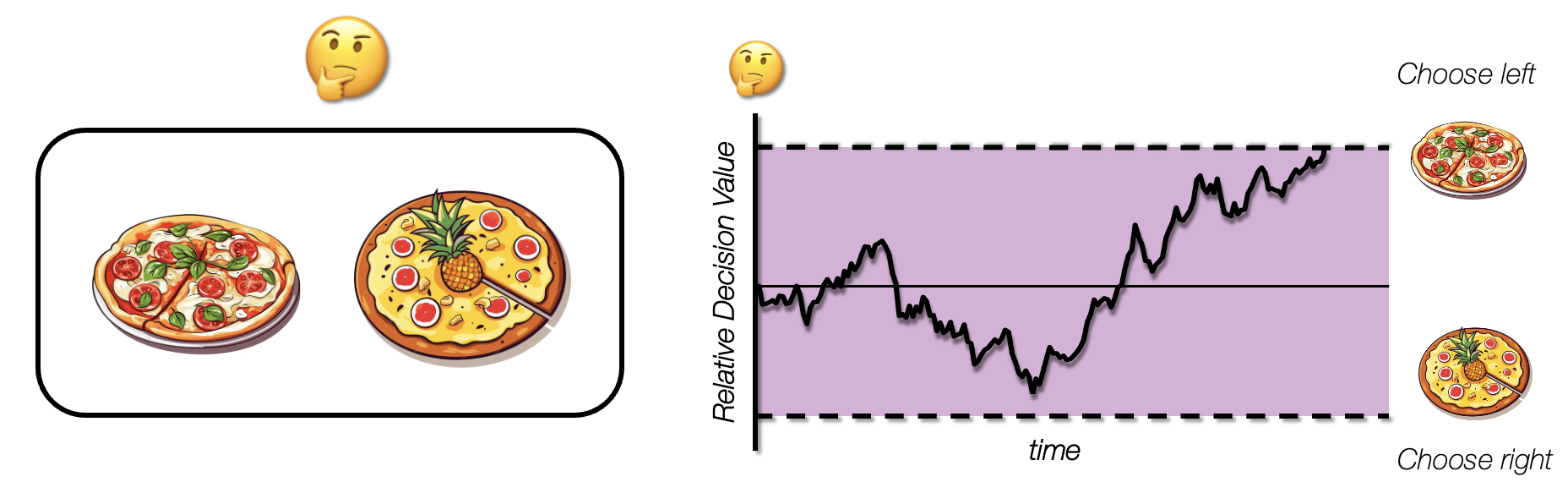}
    \caption{An example of applying the sequential sampling modeling framework to a choice between two pizzas. The decision-maker samples evidence in favor of both options until reaching a decision boundary.}
    \label{fig:exampleSSM}
\end{figure}

\section{SequentialSamplingModels.jl}

While existing methods can already simulate a wide range of SSMs across multiple languages, developments in Julia have either not been established or lack maintenance. SequentialSamplingModels (SSM.jl) provides the first unified interface for simulating validated SSMs, drawn from the literature to represent those most widely used. The package offers a user-friendly method for simulating a range of SSMs of interest to researchers. It is the first attempt to integrate a package into the Julia ecosystem for simulating and evaluating a popular model of decision-making.

\section{Availability, Development, and Documentation}

SequentialSamplingModels.jl is available on the Julia package registry. The package is maintained on GitHub at \url{https://github.com/itsdfish/SequentialSamplingModels.jl}, and the documentation can be found at \url{https://itsdfish.github.io/SequentialSamplingModels.jl/dev/}. For a more comprehensive exploration of the package functionalities and further details on future developments, the user is invited to consult the package README, documentation, and issues.

\section{Example one: simulating choices and response times}

SequentialSamplingModels.jl is an extension of the established functionalities within the Distributions package \cite{dahua_lin_2024_13838344}, designed for generating multivariate choice-RT distributions. To achieve this, a new abstract type was defined to support both continuous and discrete values. Then, a suite of constructors were designed to accommodate multivariate distributions with mixed support, which are necessary for generating discrete choices and continuous response times. Across the defined sub-types, distributions utilizing SSM.jl will output a NamedTuple consisting of a vector of choices and RTs. With samplers designed to integrate with Distributions.jl in mind, the constructors can leverage many of the related functions for probabilistic distributions, including sampling and likelihood evaluation. In Code Block 1, using the Diffusion Decision Model (\cite{ratcliff2016diffusion}; DDM), we provide an example of these features for simulating choices and response times from a pre-defined set of parameters, as well as computing the log-likelihood of the data.
\begin{lstlisting}[
    language = Julia, 
    numbers=left, 
    label={lst:example1},  % First example
    caption={A code example illustrating how to create a Diffusion Decision Model object, generate simulated data, and evaluate the log likelihood of the simulated data.}
]
using SequentialSamplingModels
# Create DDM distribution
dist = DDM(; ν = 1.00, α = 0.80, τ = 0.30, z = 0.50)
# Sample 10,000 simulated data from the DDM
sim_data = rand(dist, 10_000)
# compute log likelihood of simulated data
logpdf(dist, sim_data)

\end{lstlisting}

\begin{figure}[!htp]
\includegraphics[height=200px,width=200px]{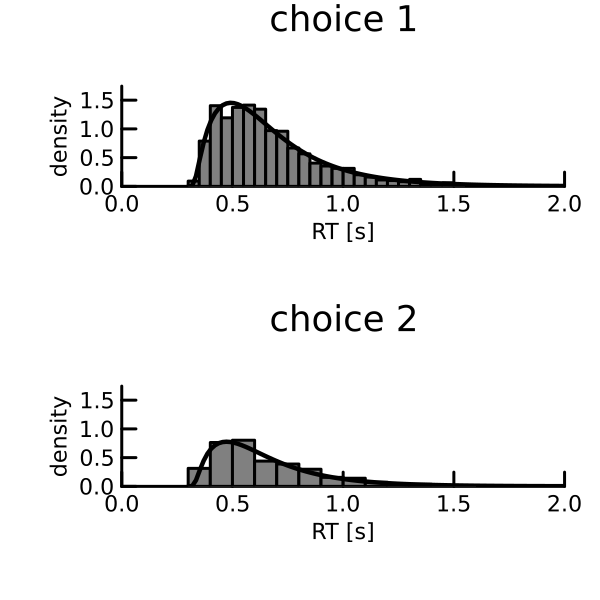}
 \caption{RT distributions of the racing diffusion model based on Code Block~\ref{lst:example2}} 
 \label{fig:rdm_choice_rt_plot}
\end{figure}


\section{Example two: plotting RT-choice distributions and simulated traces}

\begin{figure}[!htp]
\includegraphics[height=200px,width=200px]{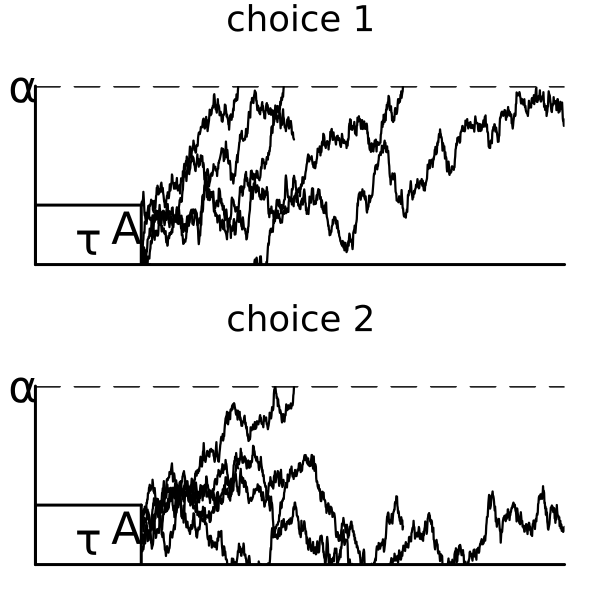}
 \caption{Five traces of the racing diffusion model based on Code Block~\ref{lst:example2}. On each simulations, the accumulators race independently until the evidence of the fastest accumulator reaches its threshold (horizontal, dashed line)} 
 \label{fig:rdm_trace_plot}
\end{figure}

SSM.jl also enables plotting functionality for SSMs when \verb|Plots.jl| (\cite{PlotsJL}) is active in your Julia session. In Code Block 2, using the racing diffusion model (\cite{tillman2020sequential}; RDM), we demonstrate how to generate two plots commonly used with SSMs. The first plot is a histogram of joint choice-RT reaction distributions paneled by choice with a probability density overlay. The output of this example is shown in Figure~\ref{fig:rdm_choice_rt_plot}. The second plot illustrates five simulated trajectories of the latent evidence accumulation process of the RDM, which is shown in Figure~\ref{fig:rdm_trace_plot}. In addition, users can animate the evidence accumulation trace with the function \verb|animate| to illustrate the model's dynamics.

\begin{lstlisting}[
    language = Julia, 
    numbers=left, 
    label={lst:example2},  
    caption={SSM.jl works with Plots.jl. The first example is a histogram of joint choice-RTs (see Figure~\ref{fig:rdm_choice_rt_plot}), and the second example is a trace of the latent evidence accumulation dynamics (see Figure~\ref{fig:rdm_trace_plot}).}
]
using SequentialSamplingModels
using Plots
# Define parameters
ν = [2,1]
k = 0.50
A = 1.0
τ = 0.30
# Create RDM distribution
dist = RDM(;ν, k, A, τ)
# Plot distribution
histogram(dist; xlims=(0, 2))
plot!(dist; t_range=range(.301, 2.5, length=100))
# Plot model simulations traces
plot_model(dist, n_sim=5, 
t_range=range(.20, 1.5, length=100), xlims=(0, 1.5))

\end{lstlisting}

\section{Example three: Bayesian inference}

In cases where closed-form likelihoods are available, SSM.jl can also facilitate parameter estimation routines available with Turing, the general-purpose probabilistic programming library (\cite{ge2018t}). Below, we provide an example of writing a Turing model using a SSM constructor and conducting a simple parameter estimation exercise
using the linear ballistic accumulator (\cite{brown2008simplest}; LBA).

\begin{lstlisting}[
    language = Julia, 
    numbers=left, 
    label={lst:example3},  
    caption={Use Turing.jl to conduct paramter estimation with a SequentialSamplingModels.jl constructor}
]
using LinearAlgebra
using Random
using SequentialSamplingModels
using StatsPlots
using Turing

Random.seed!(104)
# Simulate data
dist = LBA(ν=[3.0, 2.0], A=.8, k=.2, τ=.3)
data = rand(dist, 100)
# Define model using Turing macro
@model function model_lba(data)
    ν ~ MvNormal(zeros(2), I(2))
    A ~ truncated(Normal(.8, .4), 0.0, Inf)
    k ~ truncated(Normal(.2, .2), 0.0, Inf)
    τ ~ Uniform(0.0, minimum(data.rt))
    data ~ LBA(;ν, A, k, τ)
end
# Fit model
chains = sample(model_lba(data), NUTS(1000, .85), 
MCMCThreads(), 1000, 4)
# Examine convergence quality
summarystats(chains)
# Plot mcmc trace and posterior distributions
plot(chains)
\end{lstlisting}

In addition, SSM.jl integrates with Pigeons.jl (\citep{surjanovic2023pigeons}) and ParetoSmooth.jl (\cite{ParetoSmooth.jl}), via Turing, allowing the user to estimate the marginal likelihood and approximate leave-one-out cross-validation for Bayesian model comparison (\cite{Vehtari2017}).

\begin{figure}[!htp]
\includegraphics[height=300px,width=220px]{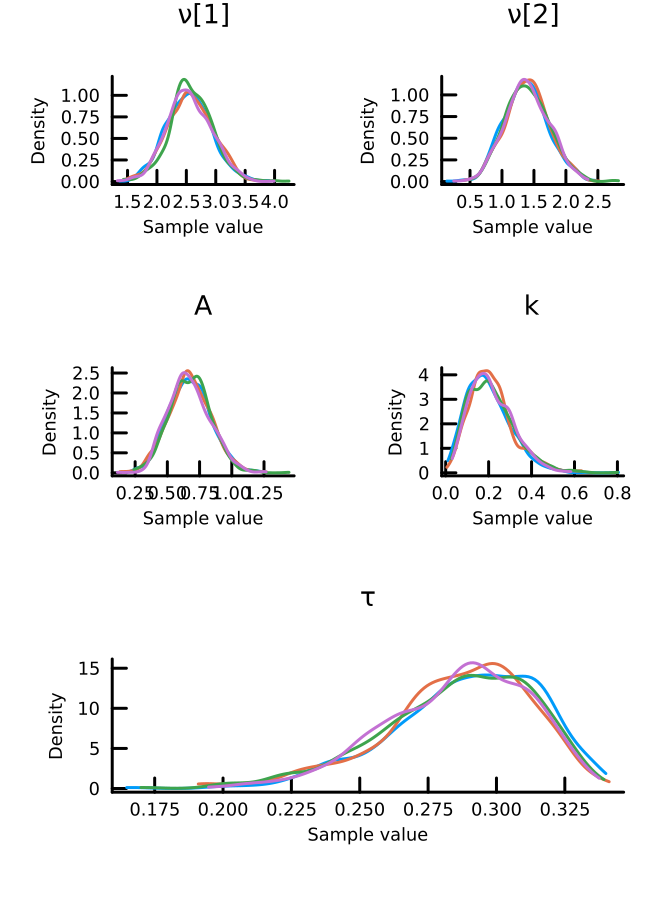}
 \caption{Posterior distributions for the parameters of the Linear Ballistic Accumulator model based on  Block~\ref{lst:example3}. MCMC trace plots were removed due to space limitations.} 
 \label{fig:lba_posterior}
\end{figure}
\section{Future Applications}

A major application of SSMs is to fit models to data via parameter estimation routines. However, not all SSMs have closed-form likelihoods that can be computed efficiently, and thus simulation-based inference has recently been proposed as a solution to accommodate parameter estimation for SSMs.

In the future, we aim to incorporate approximate solutions for likelihoods with no known closed form using approximate Bayesian computation (\cite{palestro2018likelihood}) and deep learning methods for efficient simulation-based inference, such as likelihood approximation networks (\cite{Fengler2020LikelihoodAN}) specified in Flux.jl (\cite{innesFlux}). This approach facilitates sampling from the posterior distribution of models without a closed-form likelihood function while maintaining a consistent syntax for end-users.


\bibliography{ref}

\end{document}